\documentclass[12pt]{amsart}

\newcommand{\R}{\mathbb R}
\newcommand{\x}{\mathbf x}
\newcommand{\p}{\mathbf p}

\newcommand{\reg}{\mathop{\rm reg}\nolimits}
\newcommand{\const}{const}

\begin{document}

\title
{No-Counterterm approach to quantum field theory}
\author{A. V. Stoyanovsky}
\thanks{Partially supported by the grant RFBR 10-01-00536.}
\email{alexander.stoyanovsky@gmail.com}
\address{Russian State University of Humanities}

\begin{abstract}
We give a conjectural way for computing the $S$-matrix and the
correlation functions in quantum field theory beyond perturbation
theory. The basic idea seems universal and naively simple: to
compute the physical quantities one should consider the functional
differential Schrodinger equation (without normal orderings!),
regularize it, consider the regularized evolution operator in the
Fock space from $t=T_1$ to $t=T_2$, where the interval $(T_1,T_2)$
contains the support of the interaction cutoff function, remove
regularization (without adding counterterms!), and tend the
interaction cutoff function to a constant.

We call this approach to QFT the No-Counterterm approach.
We show how to compute the No-Counterterm perturbation series
for the $\varphi^4$ model in $\R^{d+1}$. We give rough estimates
which show that some summands of this perturbation series are finite
without renormalization (in particular, one-loop integrals for $d=3$ and
all integrals for $d\ge 6$).

\end{abstract}
\maketitle

\section{The main conjecture}
In this paper we propose a conjectural way for exact computing of
the $S$-matrix and the Green functions of quantum field theory.
Recall that the Schrodinger functional differential equation reads
\begin{equation}
ih\frac{\partial\Psi}{\partial t}=\hat H(t)\Psi,
\end{equation}
where $\Psi=\Psi(t,\varphi(\cdot))$ is the unknown ``half-form''
on the space of functions $\varphi(\x)$, $\x=(x_1,\ldots,x_d)$,
$\hat H(t)=H(t,\hat\varphi(\cdot),\hat\pi(\cdot))$ is the quantum
Hamiltonian of the theory, and the operators
$\hat\varphi(\x)=\varphi(\x)$ and
$\hat\pi(\x)=-ih\frac{\delta}{\delta\varphi(\x)}$ satisfy the
canonical commutation relations
\begin{equation}
[\hat\varphi(\x),\hat\pi(\x')]=ih\delta(\x-\x'),\ \
[\hat\varphi(\x),\hat\varphi(\x')]=[\hat\pi(\x),\hat\pi(\x')]=0.
\end{equation}
For the relativistically invariant generalization of the
functional differential Schrodinger equation, see [1]. For
example, for the scalar field with self-action in $\R^{d+1}$ the
Hamiltonian reads
\begin{equation}
\begin{aligned}{}
H(t,\varphi(\cdot),\pi(\cdot))&=H_0(\varphi(\cdot),\pi(\cdot))\\
&+\int\left(\frac1{k!}g(t,\x)\varphi(\x)^k+j(t,\x)\varphi(\x)\right)d\x,\\
H_0(\varphi(\cdot),\pi(\cdot))&=\int\frac12\left(\pi(\x)^2+\sum_{j=1}^d\varphi_{x_j}(\x)^2+m^2\varphi(\x)^2\right)d\x.
\end{aligned}
\end{equation}
Here $g(t,\x)$ (the interaction cutoff function) and $j(t,\x)$
(the source) are smooth functions with compact support. For
simplicity of exposition, below we restrict ourselves by this
model. (One can see that equation (1,3) has no nonzero solutions
if $\Psi$ is a usual functional of $\varphi(\x)$, see [1].)

Let us regularize the operators $\hat\pi(\x)$ and
$\hat\varphi(\x)$, as in [2], as follows: consider the delta-like
family of smooth functions with compact support
$f_\Lambda(\x)\to\delta(\x)$, where $\Lambda\to\infty$ is the
regularization parameter (the ultraviolet regularization at small
distances), and a family of smooth functions with increasing
compact support $g_L(\x)\to1$ as $L\to\infty$ (the infrared regularization
at big distances), and put
\begin{equation}
\begin{aligned}{}
\hat\varphi_{\Lambda,L}(\x)&=g_L(\x)\int f_\Lambda(\x-\x_1)\hat\varphi(\x_1)d\x_1,\\
\hat\pi_{\Lambda,L}(\x)&=g_L(\x)\int
f_\Lambda(\x-\x_1)\hat\pi(\x_1)d\x_1.
\end{aligned}
\end{equation}
Consider the regularized Schrodinger functional differential
equation
\begin{equation}
ih\frac{\partial\Psi}{\partial t}=\hat H^{\Lambda,L}(t)\Psi,
\end{equation}
where
\begin{equation}
\hat
H^{\Lambda,L}(t)=H(t,\hat\varphi_{\Lambda,L}(\cdot),\hat\pi_{\Lambda,L}(\cdot)).
\end{equation}
The regularized quantum Hamiltonian $\hat H^{\Lambda,L}(t)$ and the
regularized free quantum Hamiltonian
\begin{equation}
\hat
H_0^{\Lambda,L}=H_0(\hat\varphi_{\Lambda,L}(\cdot),\hat\pi_{\Lambda,L}(\cdot))
\end{equation}
are well-defined and regular operators in the Fock Hilbert space
of functionals
$$
\Psi(\varphi(\cdot))=\Psi_0(\varphi(\cdot))\exp\left(-\frac1{2h}\int\tilde\varphi(\p)\tilde\varphi(-\p)\omega_\p
d\p\right),
$$
where $\p=(p_1,\ldots,p_d)$,
$\tilde\varphi(\p)=\frac1{(2\pi)^{n/2}}\int
e^{-i\p\x}\varphi(\x)d\x$, $\omega_\p=\sqrt{\p^2+m^2}$.

Denote by $U_{\Lambda,L}(T_1,T_2)$ the evolution unitary operator of
equation (5) in the Fock space from $t=T_1$ to $t=T_2$, and choose
the numbers $-T_1,T_2$ so large that the supports of the functions
$g(t,\x)$ and $j(t,\x)$ be contained in the interval $(T_1,T_2)$.
Denote
\begin{equation}
S_{\Lambda,L}(g(\cdot),j(\cdot))=e^{iT_2\hat
H_0^{\Lambda,L}/h}U_{\Lambda,L}(T_1,T_2)e^{-iT_1\hat H_0^{\Lambda,L}/h}.
\end{equation}
Clearly, this unitary operator in the Fock space does not depend
on $T_1,T_2$.

{\bf The Main Conjecture.} {\it The strong limit
\begin{equation}
S(g(\cdot),j(\cdot))=\lim\limits_{\Lambda,L\to\infty}S_{\Lambda,L}(g(\cdot),j(\cdot))
\end{equation}
is correctly defined modulo multiplication by a phase factor
$e^{ic}$, for $c$ a real number, and does not depend on the way of
regularization \emph(i.~e., on the choice of the functions
$f_\Lambda(\x)$, $g_L(\x)$\emph). The strong limit
\begin{equation}
S(g,j(\cdot))=\lim\limits_{g(t,\x)\to g}S(g(\cdot),j(\cdot))
\end{equation}
exists and coincides with the generating functional for the
operator Green functions, and the unitary operator
\begin{equation}
S(g)=S(g,j\equiv 0)
\end{equation}
coincides with the physical $S$-matrix. }

This Conjecture is partly a mathematical conjecture, and partly a
conjectural physical law.

\section{Discussion}

In this Section we present heuristic arguments in favor of the
Main Conjecture from \S1, and discuss the mathematical and
physical contents of this Conjecture.

Regarding the mathematical contents of the Conjecture, one can
imagine that there exists a space of distribution ``half-forms''
(or ``half-densities'') $\Psi$ on the Schwartz space of functions
$\varphi(\x)$, and that there exists a mathematical theory of
functional differential equations (for example, like the
Schrodinger functional differential equation above) with solutions
in this space of half-forms. (It was the main aim of Dirac in his
book [3] to construct a similar space for fermions.) The Fock
spaces are parts of this space of half-forms. Then a surprising
and mysterious fact which follows from the physical picture and
which I do not understand, is that the result of evolution of the
Schrodinger functional differential equation with the initial
conditions in the Fock space at $t=T_1$, returns to the Fock space
at $t=T_2$. It is clear that between $t=T_1$ and $t=T_2$ the
vector $\Psi$ leaves the Fock space. In [4,5] it is conjectured
that under the evolution of the Schrodinger functional
differential equation and its relativistically invariant
generalization from the surfaces $t=const$ to curved space-like
surfaces in space-time, the Fock space evolves into a family of
Hilbert spaces parameterized by space-like surfaces, and the
generalized Schrodinger equation yields an integrable flat
connection in this family. Even for the free scalar field, it is
proved in [6] that the result of evolution of the generalized
Schrodinger equation from the surface $t=const$ to a curved
space-like surface, for $d>1$ leaves the Fock space. The fact that
the solution of the functional differential Schrodinger equation
returns to the Fock space, is mathematically confirmed by results
of the theory of complex germ of Maslov and Shvedov ([2], cf.
[7]), which state that the result of quasiclassical evolution of
the functional differential Schrodinger equation along any
classical trajectory in the phase space returns to the Fock space.

Regarding the physical contents of the Conjecture, one should
prove that for the renormalizable theories, the Taylor series of
the $S$-matrix $S(g,j(\cdot))$ at $g=0$, $j\equiv 0$ coincides
with the renormalized perturbation series for the generating
functional of operator Green functions of the theory, since these
renormalized perturbation series are well checked by experiment.
Let us sketch a plan of such a proof.

In the book [8] by Bogolubov and Shirkov, the renormalized
perturbation series for the $S$-matrix and the Green functions are
constructed as the limit as $g(t,\x)\to g=const$ of a more general
object, the renormalized perturbation series $\tilde
S(g(\cdot),j(\cdot))$ with non-constant interaction cutoff
function $g(t,\x)$. This object is almost uniquely (up to the
change of parameters $m$, $g(x)$) characterized by the properties
of unitarity, causality, Lorentz invariance, and the
correspondence principle stating that the coefficient before
$g(x)$ in $\tilde S(g(\cdot),0)$ coincides with the normally
ordered interaction Lagrangian. After taking the limit $g(x)\to
g$, the parameters are fixed uniquely by conditions on the Green
functions of the theory (e.~g., for the $\varphi^4$ theory in
$\R^{3+1}$, the condition that the two-point Green function
$G^{(2)}(p_1,p_2)$ has poles at $p_i^2=m^2$, and the four-point
one-particle irreducible Green function
$\Gamma^{(4)}(p_1,p_2,p_3,p_4)$ equals $g$ at the point
$p_1=p_2=p_3=p_4=0$).

Note that the conditions of unitarity and causality are fulfilled
for any evolution operator (or limit of evolution operators) of
unitary evolution differential equations with $g(t,\x)$, $j(t,\x)$
as coefficients. Hence our operator $S(g(\cdot),j(\cdot))$ and its
Taylor series at $0$ satisfy these conditions. Regarding Lorentz
invariance for $S(g(\cdot),j(\cdot))$, it follows from the fact
that the Schrodinger functional differential equation (and its
regularizations) admit a relativistically invariant
generalization, the generalized Schrodinger equation [1] which
forms an integrable flat connection over the family of space-like
surfaces. Finally, the correspondence principle, say, for the
$\varphi^4$ model in $\R^{3+1}$ is an easy direct computation.
Therefore, the Taylor series of our $S(g(\cdot),j(\cdot))$
coincides with one of Bogolubov $S$-matrices $\tilde
S(g(\cdot),j(\cdot))$. The remaining check of parameters as
$g(x)\to g$ should not be a difficult task. We are so sure that we
obtain the right result due to our final argument which is the
inner conceptual simplicity of the theory.

Finally, note that regarding computational part of our approach,
it yields an algorithm of computation different from the
renormalization in the Feynman diagram technique. This is seen,
for example, already on the $\varphi^4$ model (see below). However, this part
of our investigation is not finished yet, so we leave it as a
challenging problem, especially for physically interesting
theories such as Yang--Mills theory or quantum gravity.

For further problems closely related to this paper, see [9].

\section{No-Counterterm perturbation series for the $\varphi^4$ model: the setup}

Traditional perturbation series for the $\varphi^4$ model is obtained by renormalization of the expression
\begin{equation}
T\exp\int g\,:\varphi(x)^4:/4!\,dx,
\end{equation}
where dots denote the normal ordering, and $\varphi(x)$ is the free scalar field, $x\in\R^{d+1}$.
It is easy to see that the perturbation series for the $S$-matrix in the
No-Counterterm approach is obtained by developing the expression
\begin{equation}
T\exp\int g\varphi(x)^4/4! dx
\end{equation}
(without normal orderings) into power series with respect to $g$.
This means that we first regularize the operator
$$
\varphi(x)^4\to\reg\varphi(x)^4=\varphi_{\Lambda,L}(x)^4,
$$
where $\Lambda\to\infty$ is the parameter of the ultraviolet regularization at small distances (and large momenta),
and $L\to\infty$ is the parameter of the infrared regularization at large distances (and small momenta). Next, we
develop the regularized integral (13) into series over powers of $g$, and finally we omit the regularization.

To perform this procedure, note first that we have
\begin{equation}
\varphi_{\Lambda,L}(x)^4/4!=:\varphi_{\Lambda,L}(x)^4:/4!+C_{\Lambda,L}:\varphi_{\Lambda,L}(x)^2:/2+\const,
\end{equation}
where $C_{\Lambda,L}$ and $\const$ are certain divergent constants. The latter constant can be neglected, since we are
interested in the expression only modulo an overall phase factor (see \S1).
Now the regularized integral (13) can be developed into
series by usual Feynman diagram techniques, using (14) (see, for example, [8]).
The quadratic term in (14) means that we change the propagator as
follows:
\begin{equation}
\begin{aligned}{}
&\frac1{p^2-m^2+i\varepsilon}\to\frac1{p^2-m^2+i\varepsilon}
+\frac1{p^2-m^2+i\varepsilon}gC_{\Lambda,L}\frac1{p^2-m^2+i\varepsilon}\\
&+\frac1{p^2-m^2+i\varepsilon}gC_{\Lambda,L}\frac1{p^2-m^2+i\varepsilon}gC_{\Lambda,L}\frac1{p^2-m^2+i\varepsilon}+\ldots.
\end{aligned}
\end{equation}
This sum of a geometric progression converges, for $g$ small enough, to the new propagator
\begin{equation}
\frac1{p^2-m^2+i\varepsilon-gC_{\Lambda,L}}.
\end{equation}
In the next Section the regularized integrals corresponding to Feynman diagrams with this new propagator are tested to
converge as $\Lambda,L\to\infty$.

\section{Rough estimates}

We consider the ultraviolet cutoff regularization ($d=3$)
\begin{equation}
\begin{aligned}{}
&\reg f(p)=\reg f(p_0,\p)=\reg f(p_0,p_1,p_2,p_3)\\
&=0\text{ if }|p_i|\ge\Lambda\text{ for some }i, 0\le i\le3.
\end{aligned}
\end{equation}
If the mass $m>0$, then we shall not need the infrared regularization at all.

The computation shows that (for any time $t$)
\begin{equation}
\begin{aligned}{}
&C_{\Lambda,L}=6\reg\int[\varphi_-(t,\p),\varphi_+(t,\p')]d\p d\p'\\
&=\reg\int\frac{6h\delta(\p+\p')}{2\sqrt{\p^2+m^2}}d\p d\p'\sim 3 h\Lambda^2.
\end{aligned}
\end{equation}
Substituting this into the propagator, one obtains for the one-loop ``fish'' diagram the following expression:
\begin{equation}{}
\reg\int\frac1{(p^2-m^2+i\varepsilon-3 gh\Lambda^2)((k-p)^2-m^2+i\varepsilon-3 gh\Lambda^2)}dp.
\end{equation}
(Here $k$ is the sum of ingoing $4$-momenta of the diagram.)
Let us divide each of the two brackets in the denominator by $\Lambda^2$.
Then the integrand becomes $\sim1$, and the integration domain
is a $4$-cube of size $\Lambda$. Hence the whole integral is $\sim\Lambda^{-4}\Lambda^4\sim1$, and it is finite.

However, for the simplest two-loop diagram with two outgoing edges we have the integral
\begin{equation}
\begin{aligned}{}
\reg\int&\frac1{(p^2-m^2+i\varepsilon-3 gh\Lambda^2)(q^2-m^2+i\varepsilon-3 gh\Lambda^2)}\\
&\times\frac1{(k-p-q)^2-m^2+i\varepsilon-3 gh\Lambda^2}dpdq,
\end{aligned}
\end{equation}
and the same argument shows that the integral diverges as $\Lambda^{-6}\Lambda^8\sim\Lambda^2$.

\section{Conclusion}

Thus, if we believe into the No-Counterterm Conjecture, we should conclude that the estimate above is
too rough for the two-loop diagram. Otherwise, if all the estimates above are correct, we see that for $d=3$
the ``No-Count\-er\-term approach'' is not valid and requires
counterterms, as well as the traditional approach.

It seems that our estimate is correct if considered as an upper bound for the integral. As a lower bound it can be
incorrect.

For a general $\varphi^4$ diagram
in $(d+1)$-dimensional space-time, the same argument as above gives the following estimate of the diagram integral.
Denote by $E_i$ ($E_e$) the number of internal (respectively external) edges of the diagram, by $L$ the number of
independent loops, by $V$ the number of vertices. Assume $d>2$. Then the following Theorem holds:

{\bf Theorem.} {\it The integral is no greater than
$O(\Lambda^m)$, where
\begin{equation}
\begin{aligned}{}
m&=-(d-1)E_i+(d+1)L\\
&=(d+1)(L-E_i)+2E_i\\
&=(d+1)(1-V)+2E_i\ \ (\text{since }V-E_i+L=1)\\
&=(d+1)(1-V)+4V-E_e\ \ (\text{since }4V=2E_i+E_e)\\
&=d+1-(d-3)V-E_e.
\end{aligned}
\end{equation}
Therefore, if the sign of $m$ is negative, then the limit of the integral is zero.}

At least, for $d=3$, $L=1$ and for $d\ge 6$ all the diagrams seemingly converge.

\end{document}